\documentstyle[preprint,float,aps,epsfig,amsmath,rotating]{revtex}

\begin{document}
\draft 
\tightenlines

\title{Recoil alignment in muon capture on nitrogen-14}

\author{
\mbox{ T.P.~Gorringe,$^{1}$} 
\mbox{ D.P.~Corbin,$^{1}$}
\mbox{ T.J.~Stocki,$^{2}$}
\\
}

\address{%
\mbox{$^1$  University of Kentucky, Lexington, KY 40506}
\mbox{$^2$  University of British Columbia, Vancouver, B.C., Canada V6T 1Z1}
}

\date{\today}
\maketitle

\begin{abstract}
We report a measurement of
the longitudinal alignment $A_L$
of the recoil nucleus
in the $\mu^- + ^{14}$N$( 1^+ , 0 )$
$\rightarrow$ $\nu_{\mu} + ^{14}$C$( 2^+ , 7012 )$ transition.
The experiment was performed on the M9B beamline 
at the TRIUMF cyclotron 
via the measurement of the Doppler lineshape
of the  subsequent $^{14}$C$( 2^+ , 7012 )$
$\rightarrow$  $^{14}$C$( 0^+ , 0 )$ gamma-rays.
We compare our result $A_L = 0.60 \pm 0.11$
to various model calculations,
and discuss the sensitivity to the 
induced pseudoscalar coupling,
second-forbidden effects
and $2 \hbar \omega$ wavefunction admixtures.
\end{abstract}


\section{Introduction}
\label{s: introduction}

Muon capture by complex nuclei
has been used in the study
of both the dynamics of the weak interaction
and the structure of the atomic nucleus.
Generally, 
in exclusive capture 
between discrete states
one attempts to separate the effects 
of dynamics and structure
by suitable choice of physical observables
and spin-parity sequences.
Of course in practice 
these dual aspects of muon capture 
are not completely separable,
and inevitably such studies must confront
the entanglement of dynamics and structure.

Of special interest in muon capture 
is the induced pseudoscalar coupling constant $g_p$
of the nucleon's weak axial current $A_{\mu}$.
Its value is firmly predicted by chiral symmetry \cite{Go58,Be94,Fe97}
thus making its measurement
an important test of low energy QCD.
Moreover, in complex nuclei its medium modification --
due to effects that range from core polarization and exchange currents
to partial restoration of chiral symmetry \cite{De76,De94,Rh84} --
are of considerable interest.
Unfortunately, the world data on $g_p$ from exclusive capture 
on complex nuclei is rather sparse,
it comprising the study of recoil orientations in
allowed Gamow-Teller transitions 
on $^{12}$C \cite{Po74,Po77,Ro81,Ku84}
and $^{28}$Si \cite{Br95,Mo97,Br00},
the study of hyperfine effects in
allowed Gamow-Teller transitions 
on $^{11}$B \cite{De68,Wi02}
and $^{23}$Na \cite{Go94,Jo96},
and the measurement of the 
$^{16}$O$(0^+, 0) \rightarrow ^{16}$N$(0^-, 120)$ first-forbidden
transition rate \cite{De69,Ka73,Gu79}.
While the majority of these experiments
are in agreement with the chiral prediction for the induced coupling,
the result of $g_p/g_a = 1.0^{+1.1}_{-1.2}$ \cite{Go04}
from the $^{28}$Si experiments
is in stark disagreement
with the theoretical expectation.

It is perhaps surprising
-- given that fifty years have passed since the
birth of $V$-$A$ theory -- 
that few measurements of recoil orientations
in muon capture have actually been performed. 
Partly motivated by the puzzling result
for the recoil orientation
in the $^{28}$Si$( 0^+  , 0 ) 
\rightarrow  ^{28}$Al$( 1^+  , 2201 )$ experiment,
we therefore decided a further investigation
of recoil orientations 
in exclusive capture 
was worthwhile.
The case we chose was 
$^{14}$N$( 1^+ , 0 )$ $\rightarrow$ $^{14}$C$( 2^+ , 7012 )$,
it presenting an interesting example
of the different sensitivities
of the different observables
to the weak dynamics 
and the nuclear structure.

Herein we report the measurement of the recoil
alignment in the transition 
$^{14}$N$( 1^+ , 0 )$ $\rightarrow$ $^{14}$C$( 2^+ , 7012 )$.
In Sec.\ \ref{s: method} we briefly outline the 
measurement technique and
experimental setup.
The determination
of the recoil alignment
from the Doppler broadened spectrum
of the 7012~keV gamma-rays
is described in detail in Sec.\ \ref{s: alignment}.
In Sec.\ \ref{s: interpretation}  we compare our experimental result
to model calculations,
and discuss the sensitivity 
to the induced pseudoscalar coupling $g_p$,
second-forbidden contributions,
and $A = 14$ nuclear wavefunctions.
Note  that gamma-ray intensities
from this $\mu^-$$^{14}$N experiment
were published earlier
in Stocki {\it et al.}\ \cite{St02}.

\section{Doppler method and experimental setup}
\label{s: method}

The products of exclusive muon capture 
are a left-handed muon neutrino 
and an orientated recoil nucleus.
The recoil orientation is a direct manifestation
of the $V$-$A$ character of the weak interaction,
and for certain spin-parity sequences 
the induced coupling $g_p$ 
has a large effect
on the recoil orientation.
Moreover, in appropriate cases the recoil orientation
can be experimentally determined from the 
directional correlation that is imparted 
on either the beta-rays
or the gamma-rays that are subsequently emitted 
by the unstable recoil.

A method for measuring  
the $\gamma$--recoil  directional correlation 
was originally proposed 
by Grenacs {\it et al.}\ \cite{Gr68}.  
The method takes advantage 
of the  Doppler shift 
of the $\gamma$--ray energy 
for a decay in--flight.
If the recoil is in motion
as it decays its energy is shifted by
\begin{equation}
\label{e: doppler}
\frac{\Delta E }{ E_o }  = \frac{ E - E_o }{ E_o } =  \beta \cos{ \theta } 
\end{equation}
where $E_o$ is the $\gamma$--ray energy in the recoil reference frame,
$E$ is the $\gamma$--ray energy in the laboratory reference frame,
$\beta = v / c$ is the velocity of the recoil in the laboratory, and
$\theta$ is the angle between the $\gamma$--ray momentum vector 
and the recoil momentum vector.
Consequently the lineshape of the Doppler broadened $\gamma$-ray
is a reflection of the $\gamma$--recoil angular correlation.
In suitable cases, {\it i.e.}\
where the $\gamma$-ray lifetime is short
and the spin-parity sequence is sensitive,
this method thus permits the determination 
of the recoil orientation in the capture process.

In this work we have studied the sequence
$^{14}$N$( 1^+  , 0 )$ $\rightarrow$  $^{14}$C$( 2^+ , 7012 )$
$\rightarrow$  $^{14}$C$( 0^+ , 0 )$.
Because the lifetime of the 7012~keV state
is rather short, $9.0 \pm 1.4$~fs \cite{Aj91}, 
and the slowing-down time 
of the $^{14}$C recoil 
in a liquid N$_2$ target
is rather long, $\sim$0.8~ps \cite{SRIM},  
the $\gamma$-ray spectrum is Doppler broadened.
The resulting $\gamma$-recoil angular distribution
is given by \cite{Bu67,Bu70,Dm70,Ci84}
\begin{equation}
\label{e: correlation}
W ( \theta ) = 1 + A_L B_{2}  P_2 ( \cos{ \theta } )
\end{equation}
where  $P_2 ( \cos{\theta} )$ is the Legendre polynomial,
$A_L$ is the recoil's longitudinal alignment
and $B_2$ is the $\gamma$-decay correlation coefficient.\footnote{In 
Eqn.\ \ref{e: correlation} 
we omit the negligible effects of rank-4 correlations
and assume either unpolarized muonic atoms or 
perpendicular geometry (see Appendix \ref{a: correlation} 
and Ref.\ \cite{Go04} for further details).}
Note that the longitudinal alignment $A_L$ 
is a direct manifestation
of the different populations 
of the magnetic sub-states
about the recoil direction,
and is entirely governed by the capture process.
Conversely the $\gamma$-decay coefficient $B_2$
is a function of the $\gamma$-decay.
For a $2^+ \rightarrow 0^+$ transition
and a rank-2 orientation 
its value is $B_2 = - \sqrt{5 \over 14}$ \cite{Ci84}.
For further details see Appendix \ref{a: correlation}. 

A detailed discussion of the experimental setup was presented in 
Stocki {\it et al.}\ \cite{St02}
-- herein we review the features that are important
for the determination
of the recoil alignment $A_L$
in the $^{14}$N$( 1^+  , 0 ) \rightarrow  ^{14}$C$( 2^+ , 7012 )$ transition.
The experiment was conducted 
on the M9B superconducting muon channel
at the TRIUMF cyclotron. 
The negative muon beam had a momentum of 65~MeV/c 
and a flux of $2 \times 10^5$s$^{-1}$,
with an electron contamination of $\sim$20\% 
and a pion contamination of $<$0.2\%.
The incoming particles were detected 
in a three-element plastic scintillator beam telescope 
and stopped in a liquid nitrogen target.
The target was $20 \times 55$~cm in cross section
and $21$~cm in thickness with
walls of polystyrene.
Mu-metal was used to reduce the ambient magnetic field
and thereby  the muon spin precession. 
The outgoing $\gamma$-rays were detected 
at right-angles to the beam axis 
in a high-purity Ge detector 
with a Compton suppressor.
The Ge detector had an efficiency of 44\%,
an in-beam energy resolution  of 2.5~keV, 
and an in-beam time resolution of 6~ns,
for the 1.33~MeV $\gamma$-ray line of a Co-$60$ source.
The Compton suppressor comprised a
six-element annulus
of NaI(Tl) crystals.

A remark is warranted
on the hyperfine effect 
in the  $\mu^-$$^{14}$N atom.
On formation the 1S ground state 
of the $\mu^-$$^{14}$N atom
is a statistical mix 
of the two hyperfine states,
{\it i.e.} one third $F_- = 1/2$ 
and two thirds $F_+ = 3/2$. 
Hyperfine transitions, 
from the higher-lying $F_+$ state
to the lower-lying $F_-$ state,
would modify these proportions,
and consequently 
the various observables
in muon capture.
Such transitions in $\mu$-atoms
are mediated by the M1 emission 
of an Auger electron from the 
surrounding electronic orbits \cite{Cu61,Wi63}. 
For $\mu^-$$^{14}$N,
where the muonic atom hyperfine splitting  is 7.4~eV 
and outer-most electron binding energy is 11.3~eV,
one expects such transitions 
to be energetically forbidden \cite{St02}.
We note the authors of Ref.\ \cite{Is86}
have suggested the possibility  
of a non-zero hyperfine transition rate
via an unconventional transition mechanism
in the $\mu^-$$^{14}$N atom.
However, in absence of confirmation 
in recent studies
by Stocki {\it et al} \cite{St02},
we will hereafter assume
that the capture originates 
from a statistical mixture
of the hyperfine states,
{\it i.e.} the measured recoil alignment
is the statistical recoil alignment $A_L^{stat}$.

Finally we note the proximity of the interesting 7012~keV level
to the $^{14}$C neutron separation energy.
This allays the concern over
cascade feeding to the 7012~keV level
from muon capture to a higher-lying level.

\section{Alignment determination}
\label{s: alignment}

\begin{figure}
\begin{center} 
\mbox{\epsfig{figure=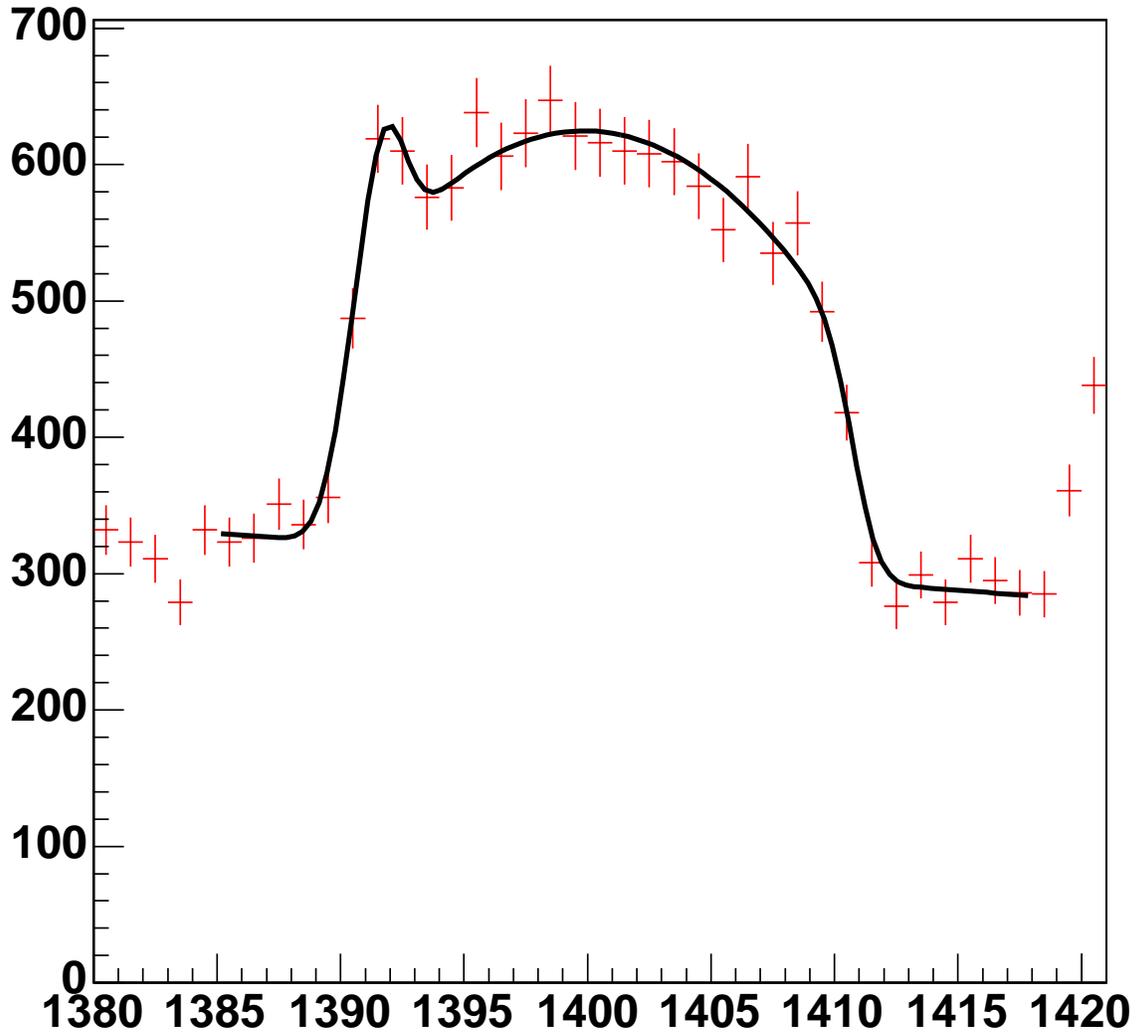,height=16.0cm}}
\end{center}
\caption{The Doppler lineshape of the 7012~keV gamma-ray
from the 
$\mu^- + ^{14}$N$( 1^+ , 0 )$ 
$\rightarrow$ $\nu_{\mu} + ^{14}$C$( 2^+ , 7012 )$
$\rightarrow$ $\gamma + ^{14}$C$( 0^+ , 0 )$ 
sequence. The rounded top of the 7012~keV lineshape is
a reflection of the 
$^{14}$C recoil alignment in the $\mu$ capture process.
The small peak on the far lefthand side of the Doppler lineshape
is the 6978~keV background line. The solid line is the
benchmark fit to the
Doppler lineshape (see text for details).}
\label{fig1}
\end{figure}

The resulting spectrum in the energy region of the 7012~keV $\gamma$-ray
is shown in Fig.\ \ref{fig1}.
The $\gamma$-ray events in Fig.\ \ref{fig1} were recorded
within a  $\sim 5.0$~$\mu$s time window of a muon arrival
and in absence of any signal in the Compton suppressor.\footnote{Note that the 
first escape peak and the second escape peak
of the 7012~keV gamma-ray were also identified,
and therefore in principle were available for 
the determination of the Doppler lineshape. 
However, in practice such analysis was precluded 
by the poor statistics and the line backgrounds
in the regions of these peaks.}
 
The first step in the analysis
was the determination of the energy calibration
and the instrumental resolution of the Ge detector.
The energy calibration was obtained from eight well-known 
and  clean $\gamma$-ray lines from thermal neutron capture on 
Al, Fe and Cl with energies that ranged from 6.2~MeV to 7.8~MeV \cite{ENSDF}. 
In fitting the peaks we found a simple Gaussian lineshape
with energy-independent width 
to be sufficient. 
A linear relation between channel number and
$\gamma$-ray energy was also adequate for the calibration
in the energy range $6.2$-$7.8$~MeV.

Next we performed the least-squares fit to the Doppler lineshape
of the 7012~keV $\gamma$-ray.
The fit function involved the
convolution of the theoretical lineshape (Eqn.\ \ref{e: correlation})
with the instrumental resolution 
of the Ge detector
and the slowing-down effects 
in the target material.
To compute the slowing-down time $\tau_{sd}$ of the $^{14}$C recoil ion
in the liquid $N_2$ target we used the computer program SRIM \cite{SRIM}.
We note that the slowing-down time $\tau_{sd} \sim 0.8$~ps
is much longer than the gamma-ray lifetime $\tau = 9.0 \pm 1.4$~fs,
and consequently in incorporating 
the slowing-down effects into the Doppler lineshape 
we assumed a simple linear form $\beta ( t ) = \beta_o ( 1 - t / \tau_{sd} )$
for the recoil velocity time dependence.

The energy spectrum in Fig.\ \ref{fig1} also indicates
the presence of a weak background line
on the far lefthand side of the Doppler spectrum.
We identified this peak as
a 6978~keV gamma-ray originating 
from thermal neutron capture
on chlorine-35.
This identification was based on both its
energy correspondence and reasonable consistency 
with observed intensities of other stronger 
$^{35}$Cl$(n,\gamma)$ lines.
The $^{35}$Cl$( n , \gamma )$ background
was believed to originate from 
polyvinyl chloride tape
in the vicinity of the target.
Consequently, both a Doppler peak and a Gaussian peak
were needed to fit the region of
the 7012~keV peak.

The fit function involved a total of
eleven adjustable parameters.
They comprised the amplitudes ($A_D$, $A_G$), centroids 
($X_G$, $X_D$) and instrumental widths ($\sigma$) of the 
Doppler peak and the Gaussian peak,
the initial velocity ($\beta$),
slowing-down time ($\tau_{sd}$)
and longitudinal alignment ($A_L^{stat}$)
of the $^{14}$C recoil,
and the amplitude and the slope
of the continuum background.
Fortunately, a number of parameters could be fixed in the fit,
for example the instrumental widths and peak centroids 
from the energy calibration,
the initial velocity from the reaction kinematics,
and the recoil slowing-down time from the SRIM calculation.
Also the amplitude of the 6978~keV $^{35}$Cl$( n , \gamma )$ line
could be estimated from the amplitudes
of the other $^{35}$Cl$( n , \gamma )$ lines 
and the available thermal neutron capture data \cite{ENSDF,Kr82}. 

Our benchmark fit, which clearly demonstrates the good agreement
between the measured spectrum and the theoretical function,
is shown in Fig.\ \ref{fig1}.
In this fit the instrumental width ($\sigma$), initial velocity  ($\beta$)
and slowing-down time  ($\tau_{sd}$) were all fixed 
at their calculated values.
The fit gave $A_L^{stat} = 0.60\pm0.11$
with a chi-squared $\chi^2_{pdf} = 0.7$.

To investigate the correlations between the
recoil alignment and the other parameters
we performed a series of fits, the results of which are 
summarized in Table \ref{t: fits}.
The Table shows
that the ``best-fit'' value
of the recoil alignment 
is only weakly dependent
on the parameters related 
to the energy calibration
and the instrumental resolution
(rows 1-4 of Table \ref{t: fits}).
In addition it shows that large changes 
in the input value of the slowing-down time,
which ranged from $\tau_{sd} = 0.4$~ps to
$\tau_{sd} \rightarrow \infty$,
had very little effect 
on  the ``best-fit'' value of the recoil alignment
(rows 5-6 of Table \ref{t: fits}).
Moreover, 
the choice of either a flat continuum background
or a linear continuum background
and the left-right margins of the fit region, 
had only minor effects on  the ``best-fit'' value 
of the recoil alignment.

By contrast the presence of the 6978~keV background line
had a significant impact on the least-squares-fit
to the recoil alignment.
The importance of including the 6978~keV line 
is demonstrated by comparing the
results of fits performed
with the 6978~keV peak included, 
which yielded $A_L^{stat} = 0.604 \pm 0.106$ (row 1),
and the 6978~keV peak omitted, 
which yielded $A_L^{stat} = 0.426 \pm 0.078$ (row 8).
In addition, we found a significant difference 
in ``best fit'' values of $A^{stat}_L$
when comparing the cases of 
either (i) varying the 6978~keV line amplitude
or (ii) fixing the 6978~keV line amplitude.
In order to fix the amplitude of the 6978~keV $^{35}$Cl$(n,\gamma)$ line 
we used the amplitudes of the nearby 6678~keV and 7414~keV 
$^{35}$Cl$(n , \gamma )$ lines
and their relative intensities taken from Krusche {\it et al} \cite{Kr82}.
The free-amplitude fit gave a recoil alignment $A_L = 0.604 \pm 0.106$ 
with a 6978~keV peak amplitude $A^G = 89\pm30$ (row 1).
By comparison the fixed-amplitude fit gave $A_L = 0.503 \pm 0.083$ 
with $A^G = 41$ (row 9), 
{\it i.e.} showing a difference of $\sim$1~$\sigma$
for the recoil alignment and a
discrepancy of $\sim$1.6$~\sigma$
for the peak amplitude.

The 1.6~$\sigma$ discrepancy between the 
estimated amplitude and the fitted amplitude
of the 6978~keV line may reflect
the statistics of the 6978~keV background line, 
the presence of an unidentified background line,
or the uncertainties in interpolating $^{35}$Cl$(n,\gamma )$ intensities.
We consider the result
of the free-amplitude fit 
to be most trustworthy.

In conclusion,
by fitting the 7012~keV lineshape 
we obtained the recoil alignment $A_L^{stat} = 0.60\pm0.11$, 
which indicates the $^{14}$C$( 2^+, 7012 )$ recoil 
is highly orientated following muon capture.
We found the alignment determination was relatively insensitive 
to the parameters related to the energy calibration,
instrumental resolution, and slowing-down effects,
but more sensitive to the presence of the
6978~keV $^{35}$Cl$(n,\gamma )$ background line.

\section{Interpretation}
\label{s: interpretation}

\subsection{Theoretical overview}
\label{s: overview}

The mass-$14$ system includes a well-known example 
of a highly suppressed allowed Gamow-Teller transition;
the $\beta$-decay that connects
the $^{14}$C$( 0^+ , 1) $ ground state
and $^{14}$N$( 1^+ , 0) $ ground state.
In lieu of g.s.-to-g.s. Gamow-Teller strength
the $^{14}$N$\rightarrow$$^{14}$C GT strength
is distributed across 
a few low-lying $^{14}$C excited states,
with most notable the $2^+$ doublet
at $7012$ and $8318$~keV \cite{Ar03}.
Interestingly, the $2^+$ doublet wavefunctions
involve both $(0p)^{-2}$ components {\em and} 
$(0p)^{-4}(1s0d)^2$ components \cite{Wa60,Cl68,Li72}, 
with configuration mixing
being indispensable
to understanding 
the partitioning
of the GT strength
within the $2^+$ doublet.

Calculations of the $\mu^- + ^{14}$N$( 1^+ , 0 )$ 
$\rightarrow$ $\nu_{\mu} + ^{14}$C$( 2^+ , 7012 )$ capture rate
have been performed by
Bukhvostov {\it et al.}\ \cite{Bu72}, 
Mukhopadhyay \cite{Mu73a,Mu73b},
Kissener {\it et al.}\ \cite{Ki73},
Desgrolard {\it et al.}\ \cite{De78,De79}
and Auerbach and Brown \cite{Au02}.
It is clearly established that a simple $(0p)^{-2}$ model 
for the $\mu^- + ^{14}N( 1^+ , 0 )$ 
$\rightarrow$ $\nu_{\mu} + ^{14}C( 2^+ , 7012 )$ transition
over predicts the rate by roughly a factor of 5-10.
As alluded to above, this discrepancy is understood 
as a result
of the importance
of the $(0p)^{-4}(1s0d)^2$ components
in the $^{14}C( 2^+ , 7012 )$ wavefunction,
such configurations decreasing
the wavefunction overlap
between the $^{14}$N initial state 
and the $^{14}$C final state.

The longitudinal alignment 
of the $^{14}$C recoil
in the $^{14}$N$( 1^+  , 0 )$$\rightarrow$$^{14}$C$( 2^+ , 7012 )$ transition
has been studied theoretically 
by Bukhvostov and Popov \cite{Bu67,Bu70}
and Dmitrieva {\it et al.}\ \cite{Dm70}.\footnote{Actually
the authors give results 
for the $\gamma$-recoil angular correlation coefficient
in the $^{14}$N$( 1^+  , 0 )$ $\rightarrow$ $^{14}$C$( 2^+ , 7012 )$
$\rightarrow$ $^{14}$C$( 0^+ , 0)$ sequence.
However, as discussed in Appendix \ref{a: correlation},
the $\gamma$-recoil correlation coefficient  $a_2$
and recoil alignment $A_L$ are 
related according to $a_2 = -\sqrt{5 \over 14} A_L$
for  $^{14}$N$( 1^+  , 0 )$ $\rightarrow$ $^{14}$C$( 2^+ , 7012 )$
$\rightarrow$ $^{14}$C$( 0^+ , 0)$.}
Bukhvostov and Popov \cite{Bu67,Bu70} 
were first to advocate the alignment 
as a possible probe
of the proton's induced pseudoscalar coupling $g_p$.
Dmitrieva {\it et al.}\ \cite{Dm70} then extended this work
to calculate both the recoil alignment
and its hyperfine dependences
using a $(0p)^2$ model space
and including the forbidden nuclear matrix elements.
For $g_p / g_a \simeq 7$ they obtained
an alignment $A^{stat}_L \simeq 0.60$
when including the second-forbidden corrections
and an alignment $A^{stat}_L \simeq 0.23$
when excluding the second-forbidden corrections, 
the authors tracing the striking contribution
of forbidden terms to the 
operator $\bf{M}_{22} \cdot \bf{\sigma}$.
The possible effects 
of the $(0p)^{-4}(1s0d)^2$ components
of the $^{14}$C$( 2^+ , 7012 )$ wavefunction
on the $^{14}$C recoil alignment 
were however not considered
in Refs.\ \cite{Bu67,Bu70,Dm70}.

\subsection{Model calculations}
\label{s: models}

Herein we have performed shell model calculations of the capture rate and the
recoil alignment for the 
$^{14}$N$( 1^+  , 0 )$$\rightarrow$$^{14}$C$( 2^+ , 7012 )$ transition
in both a $(0p)^{-2}$ model space 
and a $(0p)^{-2}$+$(0p)^{-4}(1s0d)^{2}$ model space.
For the $(0p)^{-2}$ calculation (denoted CKPOT)
we used the 0p-shell effective interaction
of Cohen and Kurath \cite{Co65}. 
For the $(0p)^{-2}$+$(0p)^{-4}(1s0d)^{2}$ calculation
we used the two-body matrix elements of: 
Cohen and Kurath \cite{Co65} for the $0p$ shell interaction,
Chung and Wildenthal \cite{Ch76} for the $1s0d$ shell interaction,
and Millener and Kurath \cite{Mi75} for the cross-shell interactions
(for further details see Warburton and Millener \cite{Wa89}).
Moreover, we performed three versions of the
$(0p)^{-2}$+$(0p)^{-4}(1s0d)^{2}$ calculation.
In version one (denoted MK3CW1) the 
$1s0d$ single particle energies
were fixed as given in Ref.\ \cite{Wa89}.
In version two (denoted MK3CW2) the 
$1s$-$0d$ single particle energies 
were shifted by $-3.0$~MeV (from Ref.\ \cite{Wa89})
in order to reproduce the $\sim$50\% admixture 
of the $(0p)^{-4}(1s0d)^2$ configurations
in the $^{14}$C$( 2^+ , 7012 )$ wavefunction
as determined in Refs.\ \cite{Wa60,Cl68,Li72}.
In version three (denoted MK3CW3) the 
$1s$-$0d$ single particle energies were 
shifted by $-4.9$~MeV (from Ref.\ \cite{Wa89})
in order to reproduce 
the $^{14}$N$( 1^+  , 0 )$$\rightarrow$$^{14}$C$( 2^+ , 7012 )$
capture rate of $\Lambda^{stat} = ( 4.4 \pm 0.6 ) \times 10^3$~s$^{-1}$
as quoted by Stocki {\it et al} \cite{St02}.

We stress our goal was not to identify one particular model
as preferable to another model.
Rather, 
by different choices of valence nucleon spaces 
and single particle energies,
we intended to expose the model
dependences of the capture rate
and the recoil alignment.
In particular, 
by varying the $1s$-$0d$ single particle energies
we changed the $(0p)^{-4}(1s0d)^2$ wavefunction admixture,
the most obvious uncertainty
in the nuclear structure 
of the $2^+$ doublet.\footnote{The dependence
of the $(0p)^{-4}(1s0d)^2$ wavefunction admixture
on the $1s$-$0d$ single particle energies
is a result 
of the proximity
of the lowest-lying $(0p)^{-2}$ 
and $(0p)^{-4}(1s0d)^2$ configurations
with spin-parity $2^+$.
This circumstance implies a large mixing
between $(0p)^{-2}$ configurations
and $(0p)^{-4}(1s0d)^2$ configurations
in the $2^+$ doublet.}
This permitted a better understanding of how the uncertainties 
in the $(0p)^{-4}(1s0d)^2$ admixture 
are mapped into the uncertainties 
in the capture rate $\Lambda^{stat}$ and the recoil alignment $A_L^{stat}$. 

\begin{figure}
\begin{center} 
\mbox{\epsfig{figure=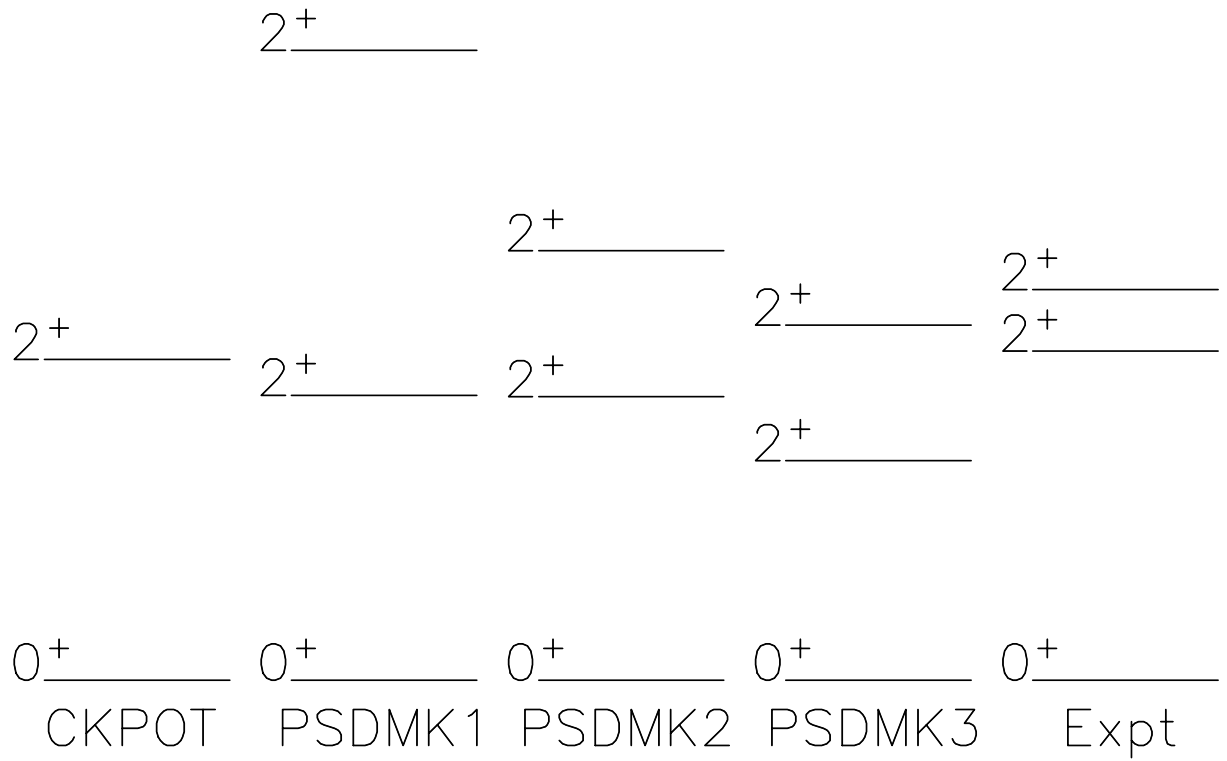,height=16.0cm,angle=90}}
\end{center}
\caption{Energy levels of the low-lying $2^+$ states
from the experimental data and the model calculations.
The experimental excitation energies of the
first and second $2^+$ states are respectively 7012 and 8318~keV.
The $2\hbar \omega$ admixture in the lowest-lying $2^+$ state 
of the three MK3CW calculations
is increasing from left to right.}
\label{fig2}
\end{figure}

A comparison of the excitation energies 
of the low-lying $(J^{\pi} , T)  = ( 2^+ , 1)$ states
from the various models with the
experimental data is given in Fig.\ \ref{fig2}.
For CKPOT the single $2^+$ state
is very nearly pure $(1p_{3/2})^7(1p_{1/2})^3$.
For MK3CW1 the lower-lying $2^+$ state
is dominantly $(1p_{3/2})^7(1p_{1/2})^3$
and the higher-lying $2^+$ state
is dominantly $(0p)^{-4}(1s0d)^2$.
Neither CKPOT nor MK3CW1
are capable of reproducing either the small energy gap 
of the 2$^+$ doublet
or the well-established mixing of the two $2^+$ states.
However, for MK3CW2 and MK3CW3,
in which we adjusted by hand the splitting
between the $0p$ shell and the $1s0d$ shell,  
the two lowest-lying $2^+$ states
are increasingly mixed,
and the energy splitting of the $2^+$ doublet 
is thus adequately reproduced.
Note
for MK3CW2 the $(0p)^{-4}(1s0d)^2$ admixture is $\sim$50\%,
this admixture dictating the choice of the single particle energies,
while for MK3CW2 the $(0p)^{-4}(1s0d)^2$ admixture is $\sim$85\%,
this admixture reproducing the capture rate 
$\Lambda^{stat} = 4.4 \times 10^3 $s$^{-1}$.
Clearly, 
the energy splitting of the $2^+$ doublet,
the $(0p)^{-4}(1s0d)^2$ content of the $2^+$ states,
and the relative positions of the 0p, 1s0d shells,
are very strongly correlated.

In order to compute the capture rate and recoil alignment 
for $^{14}$N$( 1^+  , 0 )$$\rightarrow$$^{14}$C$( 2^+ , 7012 )$
we employed the formalism of Walecka \cite{Wa75}
For completeness we give in Appendix \ref{a: calculation}
the details of our model calculations of the recoil alignment.
We took the weak nuclear current as a sum of A one-body
nucleonic currents, {\it i.e.} ignoring the effects of exchange currents,
isobar excitations, etc. 
The required nuclear matrix elements were computed
with one-body transition densities obtained
from the OXBASH shell model code \cite{Ox86}
and with harmonic oscillator radial wavefunctions
and an oscillator parameter $b = 1.7$~fm.
We set the weak vector coupling $g_v = 1$ 
and the weak magnetic coupling $g_M = 3.706$,
and  assumed the induced scalar coupling $g_s$ and 
induced tensor couplings $g_t$ were zero.
In order to scale the weak vector and magnetic couplings to finite $q^2$
we assumed a dipole form factor with $\Lambda^2$ $=$ 0.73 GeV$^2$.
The three-momentum transfer $q = 97$~MeV/c was computed
via energy-momentum conservation
\begin{equation}
\label{e: momentum}
q+q^2/2M_t = m_{\mu} - \Delta E - \epsilon_b
\end{equation}
where M$_t$ is the target mass,
$\Delta$E is the $^{14}$N$-$$^{14}$C$^{\star}$ 
nuclear binding energy difference,
and $\epsilon_b$ is the muon binding energy.
The muon wavefunction was assumed to be constant
over the nucleus,
with a reduction factor $R = 0.84$ accounting 
for the difference between a
point nucleus and a finite nucleus 
(for details see Ref.\ \cite{Wa75}). 

Lastly, to assist our later discussions of model dependences,
we make a few comments on the model calculations.
We remind the reader a 
$1^+ \rightarrow 2^+$ transition involves three multipoles
$J^{\pi} = 1^+, 2^+, 3^+$, although for
$^{14}$N$( 1^+  , 0 )$$\rightarrow$$^{14}$C$( 2^+ , 7012 )$ the contribution 
from $J^{\pi} = 3^+$ terms are very small. 
The $J^{\pi} = 1^+, 2^+$ multipoles
involve four independent amplitudes
(${\cal L}_1^5 - {\cal M}_1^5$, ${\cal T}_1^{el5} - {\cal T}_1^{mag}$,
${\cal L}_2 - {\cal M}_2$, ${\cal T}_2^{el} - {\cal T}_2^{mag5}$)
which themselves are products
of basic multipole operators
and weak coupling constants.
For $J^{\pi} = 1^+, 2^+$ the 
multipole operators are
the allowed Gamow-Teller operator $\bf{M}_{10} \cdot \bf{\sigma}$,
second-forbidden Gamow-Teller operators
$\bf{M}_{12} \cdot \bf{\sigma}$ and $\bf{M}_{22} \cdot \bf{\sigma}$,
and momentum-dependent contributions
$\bf{M}_{11} \cdot \bf{\nabla}$,
$\bf{M}_{21} \cdot \bf{\nabla}$,
$\bf{M}_{23} \cdot \bf{\nabla}$ and
$\bf{M}_{1} \bf{\sigma} \cdot \bf{\nabla}$.
The capture rate, recoil alignment and other observables 
in $^{14}$N$( 1^+  , 0 )$$\rightarrow$$^{14}$C$( 2^+ , 7012 )$ 
are all combinations of the nuclear matrix elements
of these basic multiple operators
and the weak coupling constants.
For further details see Walecka \cite{Wa75},
Donnelly and Haxton \cite{Do79}
and Appendix \ref{a: calculation}.

\subsection{Capture rate results}
\label{s: rate calc}

The $\mu^- + ^{14}$N$( 1^+ , 0 )$
$\rightarrow$ $\nu_{\mu} + ^{14}$C$( 1^+ , 7012 )$ capture rate
for a statistical mixture of the hyperfine states, 
has been measured by Babaev {\it et al.} \cite{Ba68},   
Thompson {\it et al.} \cite{Th70},
Belotti {\it et al.}\ \cite{Be76},
Giffon {\it et al.} \cite{Gi81}
and Stocki {\it et al} \cite{St02}. 
Unfortunately, the experimental results have significant scatter,
with rates 
as small as $(2.2 \pm 0.9) \times 10^3$~s$^{-1}$ \cite{St02} 
and as large as $(10 \pm 3) \times 10^3$~s$^{-1}$ \cite{Ba68}.
Stocki {\it et al} have quoted a world average of
$\Lambda^{stat} = ( 4.4 \pm 0.6 ) \times 10^3$~s$^{-1}$ \cite{St02}.

\begin{figure}
\begin{center} 
\mbox{\epsfig{figure=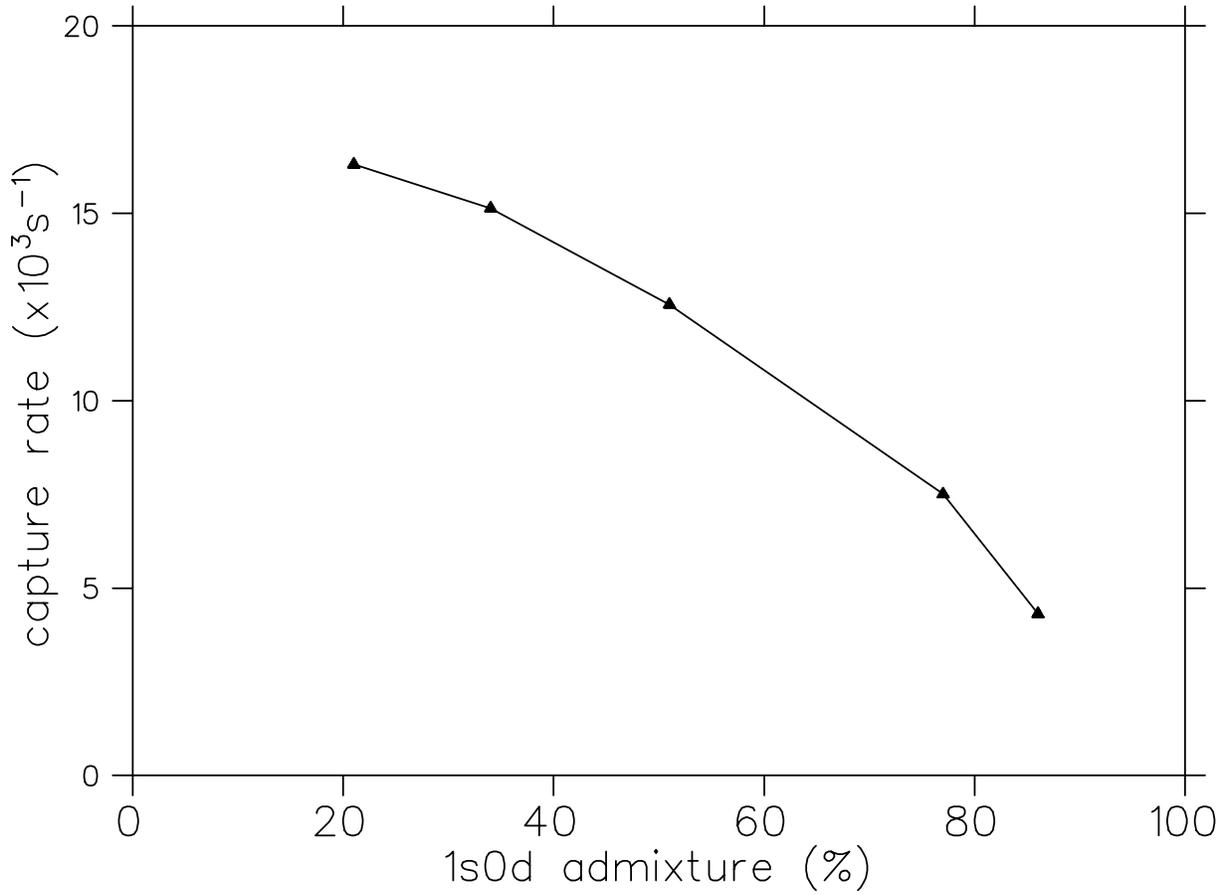,height=16.0cm,angle=90}}
\end{center}
\caption{The statistical capture rate for
$\mu^- + ^{14}$N$( 1^+ , g.s. )$ 
$\rightarrow$ $\nu_{\mu} + ^{14}$C$( 2^+ , 7012 )$
as a function of the $(0p)^{-4}(1s0d)^2$ wavefunction admixture.
The calculations were performed by varying the 
energy separation between  the $0p$ shell and the $1s0d$ shell
in the MK3CW model calculations.}
\label{fig3}
\end{figure}

Our results from the model calculations
of the muon capture rate and its hyperfine dependence
in the $^{14}$N$( 1^+  , 0 )$$\rightarrow$$^{14}$C$( 2^+ , 7012 )$ transition
are listed in Table \ref{t: rates}.
Also given in Table \ref{t: rates} are 
the $(0p)^{-4}(1s0d)^2$ admixture in the $^{14}$C final state 
and the dominant $0p_{3/2} \rightarrow 0p_{1/2}$ one-body transition density 
for the $^{14}$N$\rightarrow$$^{14}$C transition.
Note that for 
$^{14}$N$( 1^+ , 0 ) \rightarrow ^{14}$C$( 2^+ , 7012 )$
the $F = 3/2$ capture rate  and 
statistical capture rate are overwhelmingly determined 
by the product 
of the Gamow-Teller matrix element  $\bf{M}_{10}  \cdot \bf{\sigma}$
and the weak axial coupling $g_a$.
Moreover, as shown in Fig.\ \ref{fig3}, 
the capture rate  
is a strong function
of the $(0p)^{-4}(1s0d)^2$ content
of the $^{14}$C nuclear wavefunction,
the rate reflecting the decreasing wavefunction overlap 
with increasing $(0p)^{-4}(1s0d)^2$ admixture. 

As found in earlier investigations the simple $0p$-shell
grossly over-estimates the
$^{14}$N$( 1^+ , 0 )$ $\rightarrow$ $^{14}$C$( 1^+ , 7012 )$ capture rate.
Clearly, increasing the $2$$\hbar$$\omega$ admixture
by simply changing the $0p$, $1s$-$0d$ shell splitting
is capable of eliminating the  discrepancy between 
experiment and theory -- although this procedure 
is rather ad hoc.

\subsection{Recoil alignment results}
\label{s: alignment calc}

\begin{figure}
\begin{center} 
\mbox{\epsfig{figure=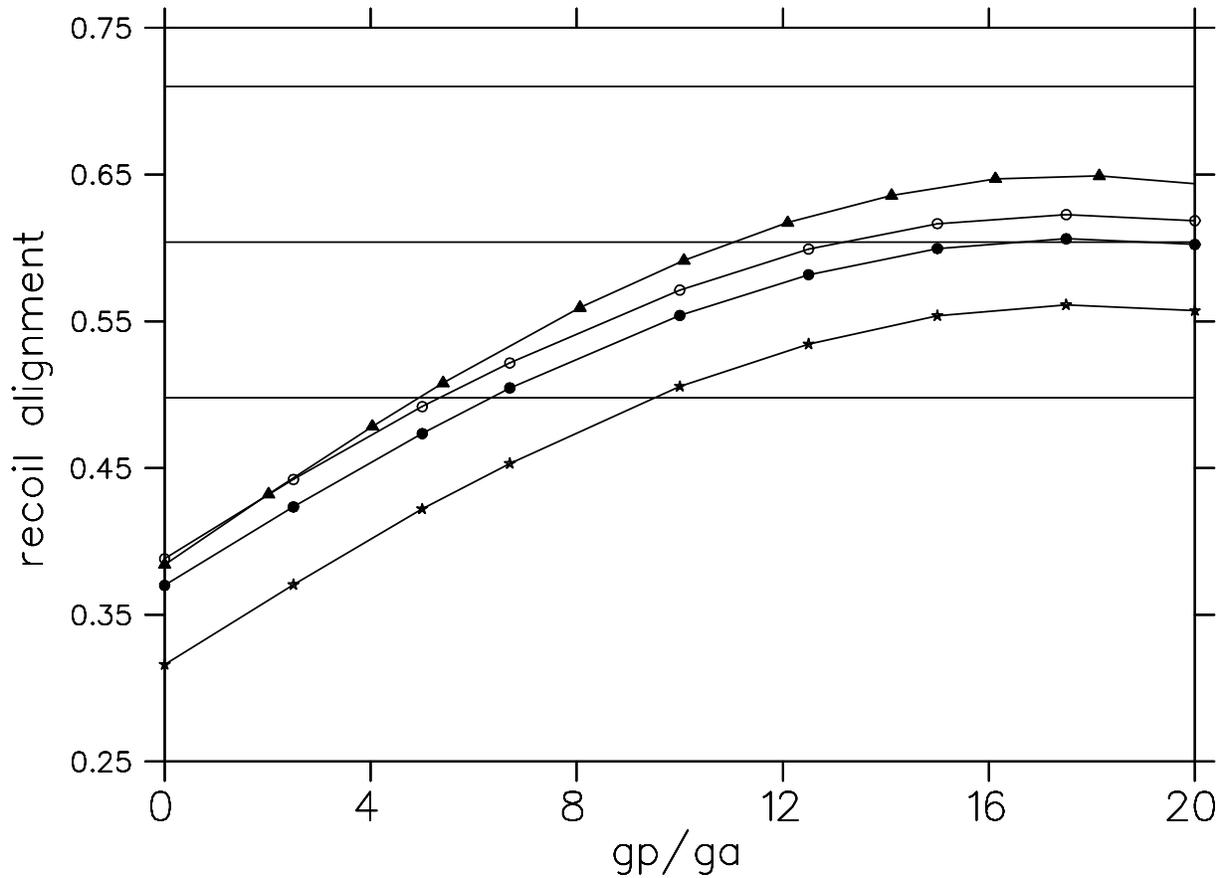,height=16.0cm,angle=90}}
\end{center}
\caption{The recoil alignment of the $^{14}$C recoil
as a function of the coupling constant ratio $g_p / g_a$.
The calculations were performed for CKPOT (diamonds),
MK3CW1 (open circles), MK3CW2 (filled circles),
and MK3CW3 (stars). The horizontal lines indicate
the central value and error bar 
for the experimental result $A_L^{stat} = 0.60 \pm 0.11$.}
\label{fig4}
\end{figure}

Our results for the model calculations of the recoil alignment 
in the $^{14}$N$( 1^+  , 0 )$ $\rightarrow$ $^{14}$C$( 2^+ , 7012 )$ transition
are plotted in Fig.\ \ref{fig4}.
Note that Fig.\ \ref{fig4} shows the recoil alignment
for a statistical mixture of the hyperfine states,
as determined by experiment and discussed in Sec.\ \ref{s: alignment}.
The alignment $A^{stat}_L$ of the statistical mixture
was obtained from the two alignments $A^{\pm}_L$ of the
hyperfine states via
\begin{equation}
\label{e: weight}
A^{stat}_L = { { f_+  A_L^+ + f_- A_L^- R } \over
{ f_+    + f_- R  } }
\end{equation}
where $f_+$, $f_-$ are the statistical
populations of the hyperfine states
and $R = \Lambda_- / \Lambda_+$ is
the hyperfine dependence of the capture rate.
Therefore, in principle, 
the alignment $A^{stat}_L$ 
is a function of the hyperfine alignments $A^{\pm}_L$
and the capture rates $\Lambda^{\pm}$,
although, in practice, 
as $R$ is small one finds 
the alignment is approximately $A_L^{stat} \simeq A_L^+$.

A useful comparison for the model calculations of the
recoil alignment is the Fujii-Primakoff approximation \cite{Mo60}.
In the Fujii-Primakoff approximation
the $\bf{M}_{10} \cdot \bf{\sigma}$ matrix element is retained
while the forbidden matrix elements are omitted.
The approximation yields 
\begin{equation}
\label{e: Ap}
A^{+}_L = +\sqrt{14}
( -X^2 -X +2 ) /
( +4  +2X +5 ) 
\end{equation}
\begin{equation}
\label{e: Am}
A^{-}_L = -{\sqrt{ 7 \over 10}}
\end{equation}
where the quantity $X = ( G_A - G_P ) / G_A $ 
and the effective coupling constants
are  $G_A = - ( g_a + ( { q }/{ 2M } ) ( g_v + g_m ) )$ and 
$G_P = - ( { q }/{ 2M } ) ( g_p - g_a + g_v + g_m )$. 
With $g_v = 1.00$, $g_m = 3.706$ and $q \sim 100$~MeV/c
one obtains a value of $A^{+}_L \simeq +0.12$ for $g_p / g_a \simeq 0$ 
and $A^{+}_L \simeq +0.35$ for $g_p / g_a \simeq 7$,
thus indicating the importance 
of the induced coupling on the recoil alignment.

For $g_p / g_a = 6.7$ the model results of Fig.\ \ref{fig4} give 
recoil alignments that range from $A^{stat}_L = +0.45$ (MK3CW3) 
to  $A^{stat}_L = +0.53$ (CKPOT).
The relatively small model dependence
for the alignment
may be contrasted 
with the relatively large model dependence 
of the capture rate,
it decreasing
by roughly
a factor of five
from CKPOT to MK3CW3
(see Table \ref{t: rates}).
To better understand this striking difference in model sensitivities
between $A^{stat}_L$ and $\Lambda^{stat}$
we remind the reader 
the alignment
is governed
by ratios 
of nuclear matrix elements
and weak coupling constants.
Therefore, 
as both the numerator and denominator in $A^{stat}_L$
are dominated by $\bf{M}_{10} \cdot \bf{\sigma}$,
the recoil alignment $A^{stat}_L$ is
rather weakly model dependent although
the matrix element $\bf{M}_{10} \cdot \bf{\sigma}$ itself
is rather strongly model dependent.
The residual model dependence of the recoil alignment $A^{stat}_L$
may be ultimately traced to the small contributions
of the $1s$-$0d$ one-body transitions densities
to the 
$^{14}$N$( 1^+  , 0 )$ $\rightarrow$ $^{14}$C$( 2^+ , 7012 )$ transition.

\begin{figure}
\begin{center} 
\mbox{\epsfig{figure=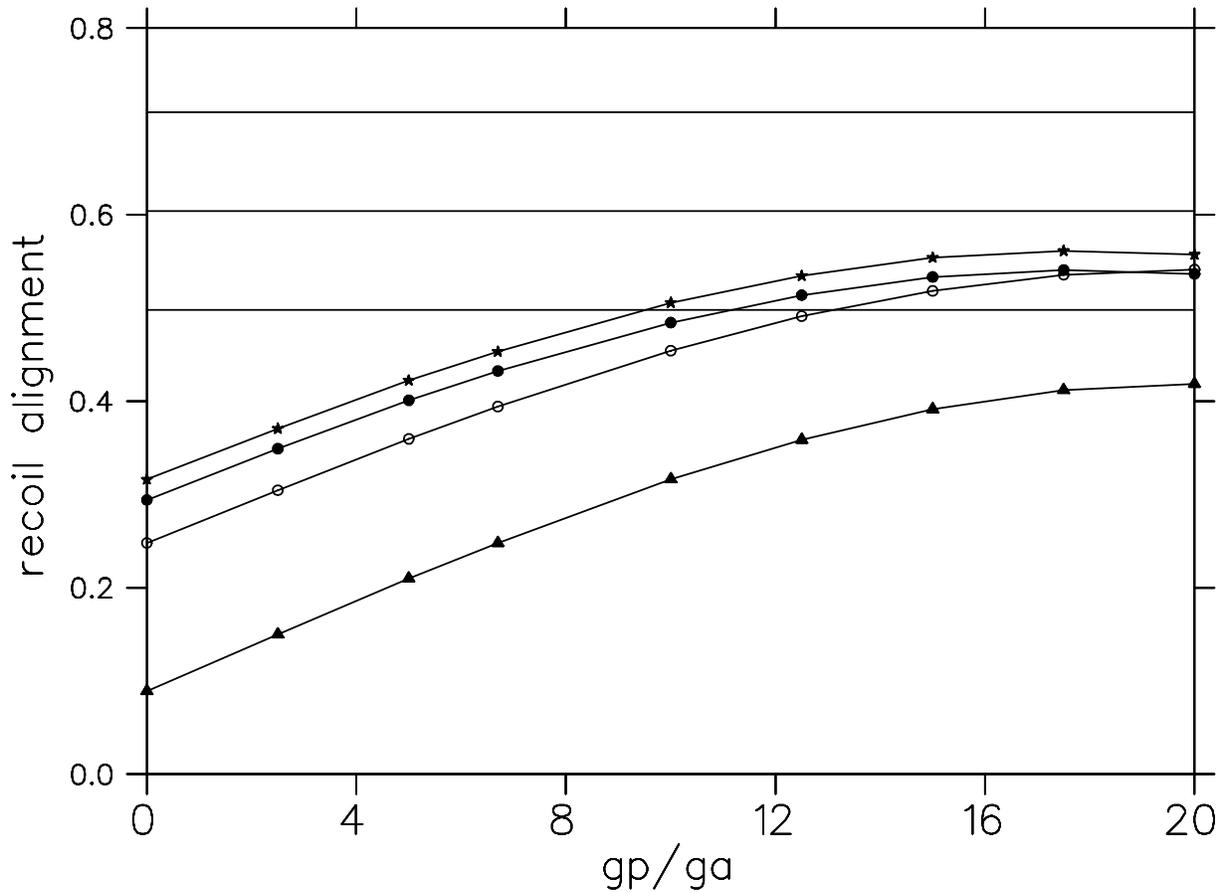,height=16.0cm,angle=90}}
\end{center}
\caption{The recoil alignment of the $^{14}$C recoil
as a function of the coupling constant ratio $g_p / g_a$.
The calculations were performed with MK3CW3 for
$M_{10} \cdot \sigma$
only  (diamonds), for  $M_{10} \cdot \sigma$ 
and  $M_{J2} \cdot \sigma$ (open circles),
for $M_{10} \cdot \sigma$, $M_{J2} \cdot \sigma$
and $M_1 \sigma \cdot \nabla$ (filled circles),
and with for $J = 1, 2, 3$ matrix elements (stars).
The horizontal lines indicate
the central value and error bar 
for the experimental result $A_L^{stat} = 0.60 \pm 0.11$.}
\label{fig5}
\end{figure}

In Fig.\ \ref{fig5} we show the relative contributions 
of the different matrix elements 
to the recoil alignment $A_L^{stat}$.
The plot reveals a substantial contribution
from the $\ell=2$ Gamow-Teller matrix elements
({\it e.g.}\  $\bf{M}_{22} \cdot \bf{\sigma}$)
and a significant contribution from the
axial charge matrix element
({\it i.e.}\ $\bf{M}_{1} \bf{\sigma} \cdot \bf{\nabla}$).
The effects of other $J = 2$ matrix elements are small and
the effects of the $J = 3$ matrix elements are negligible.
Consequently, the dominant source of model dependences 
in computing $A_L^{stat}$
is the knowledge of the ratio between
the $\ell=2$ GT matrix elements and the  $\bf{M}_{10} \cdot \bf{\sigma}$ 
matrix element. 
This ratio is sensitive to the $(0p)^{-4}(1s0d)^{2}$ wavefunction admixture
through the $1s$-$0d$ one-body transition densities.

\subsection{Coupling constant sensitivity}
\label{s: gp}

Our alignment measurement, $A^{stat}_L = 0.60 \pm 0.11$,
and alignment calculation, $A^{stat}_L = 0.45$--$0.53$,
are clearly in reasonable agreement.
Indeed the comparison between model and data 
is sufficient to constrain the coupling's value 
to $g_p / g_a \geq 5$ (CKPOT) 
to $g_p / g_a \geq 9$ (MK3CW3).
Our bound on $g_p / g_a$ is consistent 
with the results from the $^{12}$C recoil polarization experiments,
which yielded $g_p / g_a = 9.8 \pm 1.8$,
but inconsistent with the results from the $^{28}$Si recoil
alignment experiments,
which yielded $g_p / g_a = 1.0^{+1.1}_{-1.2} $.
Our result is also consistent with the prediction 
$g_p / g_a = 6.7$ of chiral symmetry arguments 
\cite{Be94,Fe97}.

In considering the dependence of $A^{stat}_L$ on $g_p$
its helpful to summarize
the terms that contribute
to $A^{stat}_L$
(for definiteness we employ the MK3CW3 model).
In the limit 
of (i) no forbidden nuclear matrix elements
and (ii) no induced pseudoscalar coupling, 
the recoil alignment is $A^{stat}_L \simeq 0.09$.
The contribution of the coupling $g_p = 6.7 g_a$ 
increases the recoil alignment to $A^{stat}_L \simeq 0.25$
and the contribution of $\ell = 2$ GT terms
increase the recoil alignment to $A^{stat}_L \simeq 0.40$.
Finally the remaining nuclear matrix elements,
mostly the axial charge matrix element,
are responsible for increasing the value to $A^{stat}_L \simeq 0.45$.  
Clearly the agreement of model and data for $A^{stat}_L$
lend support 
to significant contributions
from both the induced pseudoscalar coupling
and the $\ell=2$ forbidden corrections.

\section{Summary}
\label{s: summary}

In summary, we report a measurement of the longitudinal alignment $A^{stat}_L$
of the recoil nucleus in the 
$^{14}$N$( 1^+  , 0 ) \rightarrow  ^{14}$C$( 2^+ , 7012 )$ transition.
The alignment was determined 
from the Doppler lineshape
of the 7012~keV gamma-ray
from the $^{14}$C$( 2^+ , 7012 ) \rightarrow  ^{14}$C$( 0^+ , 0 )$ transition,
{\it i.e.}\ using the method of Grenacs {\it et al.}\ \cite{Gr68}.
Our result of $A^{stat}_L = 0.60 \pm 0.11$ indicates 
a large, positive alignment of the $^{14}$C recoil nucleus originating
from the $V$-$A$ structure of the weak interaction.

In addition, we report new calculations of the
capture rate and the recoil alignment 
for the $^{14}$N$( 1^+  , 0 ) \rightarrow  ^{14}$C$( 2^+ , 7012 )$ transition.
The calculations were conducted in both a simple $0p^{-2}$ model
space and a richer $0p^{-2} + 0p^{-4}(1s0d)^{2}$ model space.
As discussed by others,
the capture rate $\Lambda^{stat}$ is dominated 
by the $\ell = 0$ Gamow-Teller matrix element
and the weak axial coupling $g_a$,
and is highly sensitive 
to the $(0p)^{-4}(1s0d)^2$ admixture
in the $^{12}$C$(2^+ , 7012)$ wavefunction.
By comparison, 
we found
the recoil alignment $A^{stat}_L$ 
has substantial contributions 
from the induced pseudoscalar coupling 
and the second-forbidden Gamow-Teller matrix elements.
Interestingly, the recoil alignment
is comparatively insensitive
to the $(0p)^{-4}(1s0d)^2$ components
of the $^{12}$C$(2^+ , 7012)$ wavefunction.

Our measured alignment of $A^{stat}_L = 0.60 \pm 0.11$
and calculated alignments of $A^{stat}_L = 0.45$--$0.53$
are in reasonable agreement.
We view this reasonable agreement as evidence supporting
the model estimates for the
contributions of  the induced pseudoscalar coupling
and the second-forbidden corrections
to the $^{14}$C recoil alignment. 
Indeed,
by comparing the model calculations and experimental data,
we derived a bound on $g_p$ that ranges from 
$g_p / g_a \geq 9$ for the MK3CW3 calculation 
to $g_p / g_a \geq 5$ for the CKPOT calculation.
This bound is consistent 
with the value $g_p  / g_a = 9.8 \pm 1.8$ obtained
from the determination of the 
$^{12}$C$( 0^+  , 0 ) \rightarrow  ^{12}$B$( 1^+ , 0 )$ transition
recoil polarization,
but inconsistent with the value $g_p  / g_a = 1.0^{+1.1}_{-1.2}$ obtained
from the determination of the 
$^{28}$Si$( 0^+  , 0 ) \rightarrow  ^{28}$Al$( 1^+ , 2201 )$
recoil alignment.
Our result is consistent with $g_p / g_a = 6.7$ --
the theoretical prediction derived 
from chiral symmetry arguments.

Our experiment was limited
by both the statistical accuracy of the Doppler lineshape
and the unfortunate presence of the $^{35}$Cl$( n , \gamma )$ 6978~keV background line.
The model uncertainties are apparently dominated  
by the uncertainties in the $\ell = 2$ Gamow-Teller matrix elements
arising from the 
uncertainties in the $(0p)^{-4}(1s0d)^2$ wavefunction admixture.
We suggest, with better statistics and reduced backgrounds,
the $^{14}$N$( 1^+  , 0 )$ $\rightarrow$ $^{14}$C$( 2^+ , 7012 )$ transition
would offer a competitive window on induced currents and forbidden 
contributions in muon capture.

We wish to thank Drs.\ Ermias Gete, David Measday, Belal Moftah
and Michael Saliba for assistance in collecting the data
and the TRIUMF technical staff for the 
operation of the TRIUMF cyclotron. We also
acknowledge both the National Science
Foundation (USA) and the Natural Sciences and Engineering Research Council
(Canada) for their financial support.

\newpage

\appendix

\section{Relation between the $\gamma$-$\nu$ correlation
and the longitudinal alignment}
\label{a: correlation}

Herein we consider the sequence 
$\mu^- + ^{14}$N$( 1^+ , 0 ) 
\rightarrow \nu_{\mu} + ^{14}$C$( 2^+ , 7012 ) \rightarrow 
\gamma + ^{14}$C$( 0^+ , 0 )$ 
which involves an allowed  
Gamow-Teller ($1^+ \rightarrow 2^+$) transition 
and pure electric quadrupole ($2^+ \rightarrow 0^+$) decay.
We denote 
the initial, intermediate and final state angular momenta by  $J_1$, $J_2$ and $J_3$,
the neutrino and photon directions by $\hat{\nu}$ and $\hat{k}$,
and the $\gamma$-ray multipolarity by $L^{\pi}$.
For either unpolarized muons, $\vec{P_{\mu}} = 0$, 
or perpendicular geometry, $\vec{P_{\mu}} \cdot \hat{k} = 0$, 
the $\gamma$-$\nu$ directional correlation is 
given by \cite{Bu67,Bu70,Dm70,Ci84}
\begin{equation}W ( \theta ) = 1 + \sum_{s} 
a_s P_s ( \cos{ \theta } )
\label{e: W}
\end{equation}
where 
$P_s ( \cos{ \theta } )$ are the Legendre polynomials,
$a_s$ are the $\gamma$-$\nu$ correlation coefficients,
and $\cos { \theta } = \hat{\nu} \cdot \hat{k}$.
In Eqn.\ \ref{e: W} the summation involves even Legendre polynomials,
and is bounded by the smaller of $2 J_2$ or $2 L$.
Thus
for $^{14}$N$( 1^+ , 0 ) 
\rightarrow ^{14}$C$( 2^+ , 7012 ) \rightarrow 
^{14}$C$( 0^+ , 0 )$ the permissible  values are $s = 2$ and $s = 4$.

The $\gamma$-$\nu$ correlation coefficients $a_s$
are functions of both the $\mu$ capture process
and $\gamma$-decay process.
More specifically the coefficients $a_s$ 
are given by
\begin{equation}
a_s = A_s B_s
\end{equation}
where $A_s$ is dictated by the
$\mu^- + ^{14}$N$( 1^+ , g.s. ) 
\rightarrow \nu_{\mu} + ^{14}$C$( 2^+ , 7012 )$
transition,
and $B_s$ is dictated by the 
$^{14}$C$( 2^+ , 7012 ) \rightarrow \gamma + ^{14}$C$( 0^+ , 0 )$
transition.
The quantity $A_s$ is the rank-$s$ orientation
of the recoil nucleus $J_2$ about the neutrino axis $\hat{\nu}$.
The quantity $B_{s}$ is 
governed by the spin-parity sequence 
$J_2 \rightarrow J_3$
and the  multipolarity $L^{\pi}$
(a useful reference that contains tabulations 
of  $B_{s}$ coefficients is
Ciechanowicz and Oziewicz \cite{Ci84}).
For $2^+ \rightarrow 0^+$ transitions
one finds that $B_2 = - \sqrt{5 \over 14}$
and $B_4 = - \sqrt{8 \over 7}$.

Our model calculations for $\mu^- + ^{14}$N$( 1^+ , 0 ) 
\rightarrow \nu_{\mu} + ^{14}$C$( 2^+ , 7012 ) \rightarrow 
\gamma + ^{14}$C$( 0^+ , 0 )$ have shown 
the role of the rank-4 orientation 
in the $\gamma$-$\nu$ correlation
may be safely ignored.
For the rank-2 orientation
we have herein denoted $A_2$ as the recoil's longitudinal alignment 
$A_L \equiv A_2$.
The relationship between the 
$\gamma$-$\nu$ directional correlation coefficient $a_2$
and the recoil longitudinal alignment $A_L$ 
is thus simply $a_2 = - \sqrt{5 \over 14} A_L$.

\section{Model calculations of the longitudinal alignment}
\label{a: calculation}

One approach in addressing
the various physical observables in nuclear muon capture 
is the helicity representation. 
In the helicity representation
one views capture
as the two-body decay
of a spin-$F$ muonic atom
into a left-handed muon neutrino 
and a spin-$J_f$ recoil nucleus.
The rank-$s$ orientation $A^F_s$
of the recoil nucleus
about the neutrino axis
is 
\begin{equation}
\label{e: alignment}
A^F_s  =  (-1)^s \sqrt{ 2 s + 1 } \sum_{\lambda} < J_f \lambda s 0 |  J_f \lambda > 
| T^F_{\lambda} |^2
~/~ \sum_{\lambda} | T^F_{\lambda} |^2 
\end{equation}
where 
$F$ is the muonic atom hyperfine state,
$\lambda$ is the recoil nucleus helicity state,
and $T^F_{\lambda}$ are the contributing helicity amplitudes 
\cite{Mo60,Wa75,Ci84}.

For $\mu^- + ^{14}$N$( 1^+ , 0 ) 
\rightarrow \nu_{\mu} + ^{14}$C$( 2^+ , 7012 )$ 
the initial hyperfine states are
$F = J_i \pm 1/2 = 1/2 , 3/2$.
Thus according to Eqn. \ref{e: alignment},
the various observables
in $F_- = 1/2$ capture are determined
by two helicity amplitudes,
{\it i.e.}\ $T^{1/2}_{-1}$, $T^{1/2}_{0}$,
and in $F_+ = 3/2$ capture are 
determined by four helicity amplitudes,
{\it i.e.}\  $T^{3/2}_{-2}$, $T^{3/2}_{-1}$, $T^{3/2}_{0}$, $T^{3/2}_{+1}$.
These six amplitudes contain
all the dependences 
on the weak couplings 
and the nuclear structure
in the $\mu^- + ^{14}$N$( 1^+ , 0 ) 
\rightarrow \nu_{\mu} + ^{14}$C$( 2^+ , 7012 )$ transition.

More convenient for model calculations than the helicity amplitudes $T^F_{\lambda}$
are the electroweak amplitudes denoted ${\cal L}_J-{\cal M}_J$
and ${\cal T}^{el}_J-{\cal T}^{mag}_J$
where ${\cal M}_J$, ${\cal L}_J$, ${\cal T}^{el}_J$ and ${\cal T}^{mag}_J$
represent the charge, longitudinal, transverse electric and transverse magnetic operators.
For $\mu^- + ^{14}$N$( 1^+ , g.s. ) 
\rightarrow \nu_{\mu} + ^{14}$C$( 2^+ , 7012 )$ 
there are six electroweak amplitudes corresponding 
to the three multipolarities $J^{\pi} = 1^+, 2^+, 3^+$
where ${\cal L}_J\!-\!{\cal M}_J$
and ${\cal T}^{el}_J\!-\!{\cal T}^{mag}_J$
are simple products of 
basic multipole operators
and weak coupling constants 
\cite{Wa75,Do79} (the specific relations between the electroweak amplitudes
and the helicity amplitudes are given in Refs.\ \cite{Ci84}).
For 
$\mu^- + ^{14}$N$( 1^+ , 0 ) 
\rightarrow \nu_{\mu} + ^{14}$C$( 2^+ , 7012 )$ the
relevant formulas 
for ${\cal L}_J\!-\!{\cal M}_J$
and ${\cal T}^{el}_J\!-\!{\cal T}^{mag}_J$ 
with multipolarities $J^{\pi} = 1^+, 2^+$
are reproduced in Table \ref{t: operators}.
They involve the allowed Gamow-Teller operator
$\bf{M}_{10} \cdot \bf{\sigma}$,
the second forbidden Gamow-Teller operators
$\bf{M}_{12} \cdot \bf{\sigma}$ and
$\bf{M}_{22} \cdot \bf{\sigma}$,
and the 
momentum-dependent operators
$\bf{M}_{1} \bf{\sigma} \cdot \bf{\nabla}$,
$\bf{M}_{11} \cdot \bf{\nabla}$,
$\bf{M}_{21} \cdot \bf{\nabla}$ and
$\bf{M}_{23} \cdot \bf{\nabla}$.
The contributions from $J^{\pi} = 3^+$ terms
to $\mu^- + ^{14}$N$( 1^+ , 0 ) \rightarrow \nu_{\mu} + ^{14}$C$( 2^+ , 7012 )$
are very small.

Finally, the required multi-particle weak matrix elements
\mbox{$<J_f||O^{J}||J_i>$},
between an initial many-body state $| J_i >$ and final many-body state $| J_f >$,
were obtained from single--particle weak matrix elements $<
\alpha^{\prime} || O^{J} || \alpha >$, between 
an initial single--particle state 
$| \alpha > \equiv | n, j, \ell >$ 
and final single--particle state 
$| \alpha^{\prime} > \equiv | n^{\prime}, j^{\prime}, \ell^{\prime} >$, 
via
\cite{Do79}
\begin{equation}
\label{e: transition densities}
< J_f || O^{J} || J_i > =  
\sum_{\alpha , \alpha^{\prime}} C(J,  \alpha , \alpha^{\prime}, J_f, J_i) 
 < \alpha' || O^{J} || \alpha >  
\end{equation}
where $O^{J}$ denotes the relevant operator
and $C( J , \alpha ,\alpha^{\prime}, J_f, J_i)$ denotes
the so-called one--body transition densities.
The one-body transition densities govern
the contributions of the various single--particle matrix
element $< \alpha^{\prime} || O^{J} || \alpha >$ to the
multi--particle matrix element $< J_f || O^{J} || J_i >$.

\newpage

\begin{table*}
\caption{Summary of fits to the 7012~keV $\gamma$-ray lineshape 
from the $^{14}$N$( 0, 1^+ )$ $\rightarrow$ 
$^{14}$C$( 7012, 2^+ )$ transition.
$A_G$ is the amplitude of the 6978~keV background peak,
$X^D$ is the position of the 7012~KeV Doppler peak,
$\sigma$ is the instrumental resolution,
and $\tau_{sd}$ is the slowing-down time.
We also distinguish between the fits with a flat continuum
background (denoted 1 par.) 
and a linear continuum background (denoted 2 par.).
The resulting ``best-fit'' values for $A_L^{stat}$
are listed in the last column.}
\label{t: fits}
\begin{center}
\begin{tabular}{ccccccc}
  & & & & & \\ 
 $A^G$ & $X^D$ & $\sigma^{G/D}$ & $\tau_{sd}$ & $\Delta E$ & Bkgd.\ & $A_L$ \\ 
 & & & & & & \\ 
\hline
 & & & & & & \\ 
  free & free & 0.82 & 0.011 & 0.00747 & 2 par.\ & $0.604\pm0.106$ \\
  free & 7012 & 0.82 & 0.011 & 0.00747 & 2 par.\ & $0.610\pm0.092$ \\
  free & free & 1.08 & 0.011 & 0.00747 & 2 par.\ & $0.617\pm0.115$ \\
  free & free & 0.66 & 0.011 & 0.00747 & 2 par.\ & $0.592\pm0.095$ \\
  free & free & 0.82 & 0.000 & 0.00747 & 2 par.\ & $0.632\pm0.104$ \\
  free & free & 0.82 & 0.022 & 0.00747 & 2 par.\ & $0.578\pm0.105$ \\
  free & free & 0.82 & 0.011 & 0.00747 & 1 par.\ & $0.605\pm0.100$ \\
  0 & free & 0.82 & 0.011 & 0.00747 & 2 par.\ & $0.426\pm0.078$ \\
  41 & free & 0.82 & 0.011 & 0.00747 & 2 par.\ & $0.503\pm0.083$ \\
  & & & & \\ 
\end{tabular}
\end{center}
\end{table*}

\newpage

\begin{table*}
\caption{Results of the model calculations for the capture rate and
its hyperfine dependence for the  $^{14}$N$( 1^+ ,0 )$ $\rightarrow$ 
$^{14}$C$( 2^+ , 7012 )$ transition.
The hyperfine capture rates are denoted $\Lambda^{\pm}$ and the
statistical capture rate is denoted $\Lambda^{stat}$. Also
given for the different models are the 
$0p_{3/2} \rightarrow 0p_{1/2}$ one-body transition density (column five)
and the $0p^{-4}1s0d^{2}$ wavefunction admixture (column six).}
\label{t: rates}
\begin{center}
\begin{tabular}{ccccccc}
 & & & & & & \\
Model & $\Lambda_+$ & $\Lambda_-$ & $\Lambda^{stat}$ & $\Lambda_- / \Lambda_+$ & 
$0p_{3/2} \rightarrow 0p_{1/2}$ & 4h-2p \\
 & & & & & & \\
\hline
 & & & & & & \\
CKPOT     & $24.8 \times 10^3$ & $1.0 \times 10^3$ & $16.9 \times 10^3$ & 0.040 & 1.100 & 0\%  \\
MK3CW1    & $23.8 \times 10^3$ & $1.3 \times 10^3$ & $16.3 \times 10^3$ & 0.053 & 1.005 & $\sim$20\% \\
MK3CW2 & $18.3 \times 10^3$ & $1.1 \times 10^3$ & $12.6 \times 10^3$ & 0.058 & 0.808 & $\sim$50\% \\
MK3CW3 & $6.3 \times 10^3$ & $0.42 \times 10^3$ & $4.3 \times 10^3$ &  0.067 & 0.362 & $\sim$85\% \\
 & & & & & \\
\end{tabular}
\end{center}
\end{table*}

\begin{table*}
\caption{Decomposition of the electroweak operators
$ {\cal L}_J\!-\!{\cal M}_J$ and ${\cal T}^{el}_J\!-\!{\cal T}^{mag}_J$
with  multipolarities $J^{\pi} = 1^+, 2^+$
into the basic multipole operators and the weak coupling constants.
The basic multipole operators
$M_J$, $\boldsymbol{M}_{JL} \cdot \boldsymbol{\sigma}$, 
$\boldsymbol{M}_{JL} \cdot \boldsymbol{\nabla}$ 
and $M_{J} ~ \boldsymbol{\sigma} \cdot \boldsymbol{\nabla}$
are discussed in the Appendix \ref{a: calculation}.}
\label{t: operators}
\begin{center}
\begin{tabular}{lll}
 & & \\
 $J^{\pi}$ & electroweak  & multipole operator \\ 
 & operator & decomposition \\
 & & \\
\hline
 & & \\
 $1^+$~~~~~~ & $L^5_1 - M^5_1$~~~~~~ & 
$i [ ( g_a + {q \over 2 M} ( g_a - g_p )) 
( \sqrt{1 \over 3} \boldsymbol{M}_{10} \cdot \boldsymbol{\sigma} 
+ \sqrt{2 \over 3} \boldsymbol{M}_{12} \cdot \boldsymbol{\sigma} ) 
+ {q \over M} g_a M_1 \boldsymbol{\sigma} \cdot \boldsymbol{\nabla} ] \tau^{\pm}$ \\
 & & \\
 $1^+$~~~~~~ & $T^{el5}_1 - T^{mag}_1$~~~~~~ & 
$i [ ( g_a - {q \over 2 M} ( g_v - g_m )) 
( \sqrt{2 \over 3} \boldsymbol{M}_{10} \cdot \boldsymbol{\sigma} 
- \sqrt{1 \over 3} \boldsymbol{M}_{12} \cdot \boldsymbol{\sigma} ) 
+ {q \over M} g_v \boldsymbol{M}_{11} \cdot \boldsymbol{\nabla} ] \tau^{\pm}$ \\
 & & \\
 $2+$~~~~~~ & $L_2 - M_2$~~~~~~  &
$- g_v M_2 \tau^{\pm}$ \\
 & & \\
 $2^+$~~~~~~ & $T^{el}_2 - T^{mag5}_2$~~~~~~ & 
$i [ ( -g_a + {q \over 2 M} ( g_v + g_m )) \boldsymbol{M}_{22} \cdot \boldsymbol{\sigma} 
+ {q \over M} g_v ( \sqrt{3 \over 5} \boldsymbol{M}_{21} \cdot \boldsymbol{\nabla} 
- \sqrt{2 \over 5} \boldsymbol{M}_{23} \cdot \boldsymbol{\nabla} ) ] \tau^{\pm}$ \\
 & & \\
\end{tabular}
\end{center}
\end{table*}



\newpage

\newpage

\newpage

\newpage

\newpage


\end{document}